\documentclass[aps,amsmath,amssymb,prl,showpacs,11pt]{revtex4}
\begin{document}
%\mark{{Genericity of singularities...}{P S Joshi}}

\title{On the genericity of spacetime singularities}
\author{Pankaj S. Joshi}
\affiliation{Tata Institute of Fundamental Research\\ 
Homi Bhabha Road\\ 
Mumbai 400005, India}

%\keywords{Spacetime singularities, gravitational collapse, cosmology}

\begin{abstract} We consider here the genericity aspects of spacetime
singularities that occur in cosmology and in gravitational collapse. 
The singularity theorems (that predict the occurrence of singularities
in general relativity) allow the singularities of gravitational collapse 
to be either visible to external observers or covered 
by an event horizon of gravity. It is shown that the visible 
singularities that develop as final states of spherical collapse are 
generic. Some consequences of this fact are discussed.
\end{abstract}

\maketitle

\section{1. Introduction}

After the advent of singularity theorems due to Hawking, Penrose 
and Geroch in the late nineteen sixties, the role of spacetime
singularities as an inevitable feature of Einstein's theory of 
gravitation became clear. Such singularities occur in cosmology and 
in gravitational collapse. They also lead to situations where the 
gravitational field becomes ultra-strong and grows without any upper 
bound. The above-mentioned theorems further prove that the singularities 
manifest themselves in terms of the incompleteness of non-spacelike 
geodesics in spacetime.

It is possible that such singularities represent the incompleteness
of the theory of general relativity itself. Further, they may be 
resolved or avoided when quantum effects near them are included in a 
more complete theory of quantum gravity. Nevertheless, there is a 
key point here. Even if the final singularity is dissolved by quantum 
gravity, what is important is the inevitable occurrence of an ultra-strong 
gravity regime, close to the location of the classical singularity, in 
cosmology as well as during the dynamical processes involved in 
gravitational collapse. This affects the physics of the universe. An 
example of such a situation is the big bang singularity of cosmology. 
Even though such  singularities may be resolved through either 
quantum gravity effects \cite{bojo}, or features such as chaotic 
initial conditions, the effects of the super ultra-dense region of 
gravity that existed in the big bang epoch profoundly influence the 
physics and subsequent evolution of the universe.  Similarly, in the 
gravitational collapse of a massive star, the same theorems predict the 
occurrence of singularities which can be either visible to external 
observers or hidden behind the event horizon giving rise to a black hole. 
In the case of a collapse process leading to a hidden or naked singularity, 
again the important issue is how the inevitably occurring super ultra-
dense region would influence the physics outside.

The above discussion does not imply the absence of singularity-free 
solutions to Einstein's equations. There are many examples of such 
solutions (see e.g. \cite{seno} in this volume). In fact, the G\"odel 
universe is a cosmological model which is geodesically complete. Also, 
Minkowski spacetime is the vacuum solution which is geodesically complete. 
For an additional discussion of singularity free models in general 
relativity, we refer to \cite{ryan}. Again, in gravitational collapse, 
one could start with a matter cloud which is initially collapsing. 
However, there could be a bounce later and the cloud may then disperse 
away, at least in principle. In such a case, no spacetime singularity 
or ultra-strong gravity region need form.

Since its derivation in the early 1950s, 
the Raychaudhuri equation \cite{akr}, has played a central role in the 
analysis of spacetime 
singularities in general relativity. Prior to the use of this 
equation to analyse collapsing and cosmological situations for the occurrence 
of singularities \cite{he}, most works on related issues had taken up 
only rather special cases with many symmetry conditions assumed on 
the underlying spacetime. Raychaudhuri, however, considered for the first 
time these aspects {\it within the framework of a general spacetime without 
any symmetry conditions}, in terms of the overall behaviour of the 
congruences of trajectories of material particles and photons propagating 
and evolving dynamically. This analysis of the congruences of non-spacelike 
curves, either geodesic or otherwise, showed how gravitational focusing 
took place in the universe, giving rise to caustics and conjugate points.
Before general singularity theorems could be constructed, however, 
another important mathematical input was needed in addition to the 
Raychaudhuri equation. This was the analysis of the causality structure 
and general global properties of a spacetime manifold. This particular 
development took place mainly in the late 1960s (for a detailed
discussion, see e.g. \cite{wald} or \cite{joshi}).   
The work of Penrose, Hawking, and Geroch then combined these two 
important features, namely the gravitational focusing effects 
due to matter and causal structure constraints following from global 
spacetime properties, to obtain the singularity theorems.

We review these developments briefly in the next Section 2. We then 
point out that the main question here is that of genericity of such spacetime
singularities, either in cosmology or in collapse situations. The singularity
theorems, while proving the existence of geodesic incompleteness, by 
themselves provide no information either on the structure and properties of 
such singularities or on the growth of curvature in their vicinity. Some of 
these issues, together with the possible avoidance of singularities, are 
considered 
in Section 3. We then discuss the genericity aspects of visible singularities 
in Section 4. The final Section 5 contains some concluding remarks.

\section{2. The occurrence of singularities}

We discuss here the occurrence of spacetime singularities in some
detail within a general spacetime framework, and point out the crucial       
role of the Raychaudhuri equation therein. The basic ideas involved in the 
proofs of the singularity theorems are reviewed, and what these theorems  
do {\it not} imply is pointed out .

We observe the universe today to very far depths in space and time.
While looking deep into space, say for example in diametrically opposite 
directions, regions with extremely distant galaxies are seen in each 
direction. Interestingly, these regions have quite similar properties in 
terms of their appearance and the homogeneity in their galactic spatial 
distribution. These regions are, however, so far away from each other that 
they have had no time to interact mutually. That is because the age 
of the universe since the big bang has not been large enough for such 
interactions to have occurred in the past. Thus, within the big bang 
framework of cosmology, a relevant question arises: how come these regions 
have such similar properties? This is one of the major puzzles of modern 
cosmology today.

This observed homogeneity and isotropy of the universe at large enough 
scales can be modelled by the Robertson-Walker geometry. The metric of the 
corresponding spacetime is given by
\begin{equation}
ds^2 = - dt^2 + R^2(t) \big[ {dr^2\over (1-kr^2)} + r^2 d{\Omega}^2 \big]. 
\end{equation}
Here $d\Omega^2 = d\theta^2 + sin^2\theta d\phi^2$  is the
metric on a two-sphere. 
The universe is assumed to be spherically symmetric here. 
There is an additional assumption that the matter content of the 
spacetime is homogeneous and isotropic. That is, the matter density is the 
same everywhere in the universe and also looks the same in {\it all} 
directions. Combining this geometry with the Einstein equations and 
solving the same, one is led to the Friedmann solution yielding a description 
of the dynamical evolution of the universe. The latter requires that the 
universe must have had a beginning at a finite time in the past. This is the 
epoch of the so called big bang singularity. The matter density as well as 
the curvatures  of spacetime diverge in the limit of approaching this 
cosmological singularity. This is where all non-spacelike geodesics are 
incomplete at a point in the past where spacetime comes to an end.

A similar occurrence of the formation of a spacetime singularity takes place
when a massive star collapses freely under the force of its own gravity after
the exhaustion of its nuclear fuel. If the mass of the star is small enough, 
the latter can stabilize as a white dwarf or a neutron star as its endstate.
However, in case the mass is much larger, say of the order of tens of solar 
masses, a continual collapse is inevitable once there are no internal 
pressures left to sustain the star. Such a scenario was considered and 
modelled by Oppenheimer and Snyder \cite{os}, when they considered a 
collapsing spherical cloud of dust. Again, according to the equations of 
general relativity, a spacetime singularity of infinite density and 
curvature forms at the centre of the collapsing cloud.

These singularities, after their discovery, were debated extensively by 
gravitation theorists. An important question that was persistently 
asked at this juncture was the following. Why should these models be taken 
so seriously when they assume so many symmetries? Perhaps such a spacetime 
singularity was arising as a result of these assumptions and could only 
occur in such a special situation. In other words, these singularities
could be just some isolated examples. After all, the Einstein equations
\begin{equation}
R_{ab} -  {1 \over 2} R g_{ab} = 8\pi T_{ab},
\end{equation}
governing the ever present force of gravity, are a complex system of
second-order, non-linear, partial differential equations admitting an
infinite space of solutions in which the models discussed above are
isolated examples. In other words, the issue was the absence of any proof 
that singularities would always occur in a general enough gravitational 
collapse when a massive star dies, or in a generic enough cosmological 
model. In fact, there was widespread belief in the 1940s and 1950s that 
such singularities would be removed both from stellar collapse and from the 
cosmological beginning of the universe (which are two very important physical 
situations), once assumptions like the dust form of matter and the
spherical symmetry of the model were relaxed and more general 
solutions to the Einstein equations were considered. This is where the work 
by Raychaudhuri on the gravitational focusing of matter or light in a 
spacetime, and that of Penrose, Hawking, and Geroch on the causal structure 
and global properties of a general spacetime, the culmination of which 
were the singularity theorems, became relevant. These theorems showed that 
spacetime singularities, such as those depicted in the above examples, 
manifested themselves in a large class of models under quite general physical 
conditions.

We now outline the basic idea and the chain of logic behind the proof 
of a typical singularity theorem, highlighting the role of the Raychaudhuri 
equation therein. There are several singularity theorems available which 
establish the non-spacelike geodesic incompleteness for a spacetime  
under different sets of physical conditions. Each of these may be
more relevant to one or the other specific physical situation, 
and may be applicable to different physical systems such as stellar collapse 
or the universe as a whole. However, the most general of 
these is the Hawking-Penrose theorem 
\cite{hawpen}, 
which is applicable to both the collapse situation and the cosmological 
scenario. Let us first briefly describe the main steps in the 
proof of this theorem. 
Using causal structure analysis, it is first shown that between certain pairs 
of events in spacetime there must exist timelike geodesics of maximal length.
However, both from causal structure analysis and from the global properties of
a spacetime manifold $M$ (which is assumed to satisfy the generic 
condition
as well as a specific energy condition), it
follows that a causal geodesic, which is complete in regard to both the future
and the past, 
must contain pairs of {\it conjugate points} where nearby null or
timelike geodesics intersect. One is then led to a 
contradiction because
the maximal geodesics mentioned above cannot contain 
any conjugate points, the existence of which would be against their 
maximality. Thus $M$ itself must have non-spacelike geodesic incompleteness.
We restate the theorem more carefully and accurately below. 
\medskip

A spacetime   $(M,g)$ cannot be 
timelike and null geodesically complete if the following are satisfied:
\bigskip

\noindent(1) $R_{ij}K^iK^j\ge0$ for all non-spacelike vectors $K^i$;

\noindent(2) the generic condition is satisfied, that is, every 
non-spacelike geodesic contains a point at which 
$K_{[i}R_{j]el[m}K_{n]}K^e K^l\ne0$, where $K$ is the tangent to the 
non-spacelike geodesic;    

\noindent(3) the chronology condition holds, and

\noindent(4) there exists in $M$  either a compact achronal set 
without edge, or a closed trapped surface, or a point $p$ such that for 
all past directed null geodesics from $p$, eventually the expansion
parameter $\theta$ must be negative.

The first condition above is an energy condition. All classical
fields have been observed to satisfy a suitable positivity of energy
requirement. The second condition is a statement that all non-spacelike
trajectories do encounter some non-zero matter or stress-energy 
density somewhere during their entire path. The third is a global
causality requirement to the effect that there are no closed timelike
curves in the spacetime. Finally, in the last condition above, 
the first or the second part corresponds to a gravitational collapse 
situation, whereas the last part corresponds to gravitational 
focussing within a cosmological framework, where $\theta$ denotes
the expansion for the congruence of non-spacelike curves. If any of 
these conditions are satisfied then the theorem can go through.

The main idea of the proof is that one shows  
the following three conditions cannot hold simultaneously: (a) every 
inextendible non-spacelike geodesic contains pairs of conjugate points, 
(b) the chronology condition holds, (c) there exists an achronal set 
$\cal S$ in spacetime such that $E^+(\cal S)$ or $E^-(\cal S)$ is 
compact.

In the above,
For any set $\cal S$, the future of the set is denoted by 
$I^+(\cal S)$, which is the set of all those points which can be connected
by future-directed timelike curves from any point in $\cal S$. The past
of a set, denoted by $I^-$ is defined similarly dually. An achronal set
is a set of spacetime events, no two of which are chronologically related. 
The set $E(S)$ here denotes the {\it edge} of $S$ which is a set
of all points $x\in S$ such that every neighbourhood of $x$ contains
$y\in I^+(x)$ and $z\in I^-(x)$, such that there is a timelike curve 
that does not meet $S$.

%% The edge of a set...is defined as..

If the above is shown then the theorem is proved, because the condition
(3) is same as (b), the condition (4) implies (c), and conditions (1) and (2) 
imply (a). First, we note that (a) and (b) imply the strong causality 
of the spacetime  (see Proposition 6.4.6 of 
\cite{he}). Strong causality is a causality condition stronger than
the chronology which rules out the occurrence of closed timelike curves,
and ensures that once a timelike path has left an event in the spacetime,
it cannot return in the arbitrary vicinity of the same.

While for further
details and definitions, we refer to
\cite{he} or
\cite{wald},
in the following we give the brief outline of the main argument
for the proof.
It can be shown that if $\cal S$ is a future trapped set, which
essentially means that no non-spacelike paths escape its own future, and if
strong causality holds on $\overline {I^+(\cal S)}$ then there exists a future
endless trip $\gamma$ such that $\gamma\subset Int D^+(E^+(\gamma))$. 
Here the sets $D^+$ denotes the future domain of dependence for a set.
Now, one defines $T=\overline{J^-(\gamma)}\cap E^+(\cal S)$, then $T$ turns out
to be past trapped and  hence there exists $\lambda$, a past endless
causal geodesic in $Int(D^-(E^-(T))$. Then one chooses a sequence $\{a_i\}$
receding into the past on $\lambda$ and a sequence $\{c_i\}$ on $\gamma$ to 
the future. The sets $J^-(c_i)\cap J^+(a_i)$ are compact and globally
hyperbolic, so there exists a maximal geodesic $\mu_i$ from $a_i$ to $c_i$
for each $i$. The intersections of $\mu_i$ with the compact set $T$ have
a limit point $p$ and a limiting causal direction. The causal geodesic $\mu$
with this direction at $p$ must have a pair of conjugate points. This is then
shown to be contradictory to the maximality property of the geodesics
stated above.

The Raychaudhuri equation plays a key role in the above argument.
It implies the occurrence of conjugate points on 
the null or timelike geodesics under consideration. 
To state this in a somewhat different way, 
let us suppose that the spacetime manifold $M$ is non-spacelike 
geodesically complete. In that case, causal structure and the causality as 
well as 
regularity assumptions imply that there must exist {\it maximal} 
non-spacelike geodesics between certain pairs of spacetime points or along 
the non-spacelike geodesics which are orthogonal either to a spacelike 
hypersurface or to a trapped spacetime surface.
However, a suitable energy condition, such as the weak energy
condition as given by $T_{ab} V^aV^b \ge 0$, together with Einsteins equations,
implies that $R_{ab} V^aV^b \ge 0$, which in turn implies from the
Raychaudhuri equation that conjugate
points {\it must} develop along these maximal length geodesic trajectories.
This contradicts the statement, dictated by causal structure requirements, 
that maximal geodesics 
cannot contain any conjugate points.

\section{3. Singularity avoidance}

It would be quite attractive if one could avoid any such 
singularity at the classical level itself. Efforts were made in
that direction after singularities had been found in the special 
cases of FRW cosmologies and the Schwarzschild solution related 
to gravitational collapse. It was believed that, by 
going to situations which did not require those exact symmetries, one 
might get rid of spacetime singularities altogether.
However, as discussed above, the singularity theorems show that 
this is not possible. Once we accept that geodesic incompleteness
represents a genuinely singular behaviour, the only way to avoid spacetime
singularities at a purely classical level within the framework of Einstein 
gravity would be to somehow break or violate one of the 
assumptions or conditions which have been used in proving these theorems.
Let us consider this possibility in some detail. The assumptions 
used in proving various singularity theorems fall mainly into three classes, 
as pointed out above. The first is a reasonable causality condition on  
spacetime; it ensures an overall well-behaved global behaviour for 
the universe. The second is some suitable energy condition 
that emphasises the positive nature of the energy density of classical matter 
fields. The final assumption is the 
occurrence of a suitable focusing effect in spacetime, as caused by 
the presence of matter fields. This occurrence is typically in the form of  
trapped surfaces present in situations of gravitational collapse or a 
suitable overall focusing in the cosmological situation.

First, consider a possible violation of the causality condition.
Note that the Einstein equations  as such do not demand
causality of spacetime as a whole. In fact, they
impose no constraints on the global topology of spacetime. 
Hence, without breaking any consistency requirements, one could consider
causality violating spacetimes and see if somehow the 
occurrence of spacetime singularities could be avoided there. 
This issue has been analysed in some detail. It turns out
that if causality is violated in any finite region of spacetime,
that in its own right causes spacetime singularities in the form of
geodesic incompleteness
\cite{tipler}
\cite{joshipla}.
These results show that the ccurrence of closed timelike lines in any
finite region of spacetime necessarily causes singularities there. 
One could of course preserve causality in the universe by disallowing
closed non-spacelike curves, but still admitting 
higher order causality violations. The latter can occur from timelike paths 
coming back to arbitrarily close neighbourhoods of an event that they had 
left. Or, a spacetime $(M,g)$ could be causal, but allowed to admit closed
timelike curves with the slightest perturbation 
of the metric. This last situation is 
described as a violation of the stable causality condition. The above results 
show that such higher order causality violations also give rise to 
spacetime singularities.
One may then conclude that violating causality is not an effective alternative
means to avoid spacetime singularities. Of course, 
as noted earlier, the G\"odel universe is geodesically complete and 
violates causality at every event of spacetime. So there 
might be a possibility that the violation of causality, introduced at every 
point of spacetime, may be able to avoid any singularity. However, such 
a universe itself would be quite `singular' in some sense. Indeed, because of 
this and other unphysical features, the G\"odel universe has never been 
regarded as a suitable model \cite{he}.

The second alternative, of violating the energy
condition, has again been
investigated in some detail
(see e.g. 
\cite{tipler}
for a discussion on this). The essential conclusion here has been the 
following. So long as this condition, even if
violated locally, holds in a spacetime 
averaged sense, the existence of
singularities in the form of non-spacelike geodesic incompleteness
would be implied. 
Only if one violates the 
energy condition for
almost all the events in the
universe making the energy density an overall negative quantity, could one
avoid any possible singularity. This does not look to be physically
achievable, at least for classical fields. So one could conclude that the
violation of the energy condition does not 
offer an effective alternative in avoiding spacetime singularities. 
In other words, local violations of the energy condition do not 
affect the occurrence of singularities. Furthermore, no physical 
mechanisms have been observed or known which may allow such a violation of the 
energy condition in the universe.

This brings us to the final and most 
important alternative, namely that of avoiding sufficient 
gravitational focusing in spacetime by breaking 
condition (4) above in the proof of the singularity theorem. This would 
amount to disallowing the formation of any 
trapped surfaces in gravitational collapse and similarly not 
having enough convergence in a cosmological scenario. Basically, if one wants 
to avoid all trapping, without violating the energy
condition, the only option would be to have very little matter and 
stress-energy 
density, so that light rays just avoid getting sufficiently focused.
This is the direction that some recent works 
appear to point to 
\cite{seno}
\cite{nkdakr}. 
Similar results were obtained earlier, showing that 
if the spacelike surfaces have sufficient matter in some non-vanishing
average sense, then {\it all} non-spacelike trajectories would be
incomplete in the past
\cite{joshiprama}
\cite{joshibull}.
For example, if the microwave background radiation is taken to be
universal and also to have a global minimum in its energy density
over a spacelike surface, that will cause the necessary convergence;
all non-spacelike trajectories will then be incomplete in the past.
Thus the avoidance of trapped surfaces or 
cosmological focusing would be one way not to have spacetime 
singularities. However, if trapped surfaces did form
in spherical or other more general gravitational collapse processes, the 
occurrence 
of singularities will be a generic phenomenon. We can elaborate this last 
point. 
Let such trapped surfaces develop as the collapsing 
system evolves dynamically. Could we generalize these conclusions for a 
non-spherically symmetric 
collapse? Are they valid at least for small perturbations 
from exact spherical symmetry. By using 
\cite{he} 
the stability of Cauchy development in general 
relativity, one can show that the formation of trapped surfaces is a stable  
feature when departures from spherical symmetry are taken into account.  The 
argument goes as follows. Consider a spherically 
symmetric collapse evolution from given initial data on a partial Cauchy 
surface $S$. Then trapped surfaces $\cal T$ are found to form in the shape 
of all spheres with $r < 2m$ in the exterior Schwarzschild geometry.  
The stability of Cauchy development then implies that, for all initial 
data sufficiently near the original data in the compact region 
$J^{+} (S) \cap J^{-} (\cal T)$, trapped surfaces must still occur.  
The curvature singularity of a spherical collapse also turns out 
to be a stable feature as implied by the singularity theorems discussed above, 
The latter show that closed trapped surfaces always imply the existence 
of a spacetime singularity under reasonable general conditions.  
In this sense, such singularities, when they occur, are really generic.

\section{4. Genericity of naked singularities in spherical collapse}

The above consideration points to a wide variety of circumstances
under which singularities develop in general relativistic cosmologies
and in many gravitational collapse processes. Singularity theorems imply
the existence of vast classes of solutions to the Einstein equations that 
must contain spacetime singularities, as characterized by the conditions
of these theorems, and of which the big bang singularity is one 
example. These theorems therefore imply that singularities must occur 
in Einstein's theory quite generically, i.e. 
under rather general physically reasonable conditions on the underlying 
spacetime. Historically, this implication considerably strengthened our 
confidence in the big bang model which is used extensively 
in cosmology today.

As mentioned earlier, the singularities are predicted to be either
visible to external observers or hidden inside the 
event horizons of gravity. This is related to the causal structure in the 
vicinity of the concerned singularity and is 
particularly relevant to realistic 
gravitational collapse scenarios and to the basics of black hole
physics. However, these 
singularity theorems provide no further information. Yet, we need more 
information on the structure of the singularities in terms of their visibility,
curvature strengths and other such aspects. 
What is therefore called for 
is a detailed investigation of the 
dynamics of gravitational collapse within the framework of Einstein's
theory. Extensive investigations 
have been made in the past couple of decades or so 
from such a perspective. These have dealt with 
dynamical gravitational collapse, mainly for 
spherical models but also in some non-spherical cases, investigating
the structure and visibility of the concerned 
singularities. Depending 
on the initial conditions and the 
types of evolution of the collapse process allowed 
by the Einstein equations, the singularities of 
collapse can be either visible or covered. See, for example, 
\cite{rev1} 
\cite{rev2}
\cite{rev3}
\cite{rev4}
\cite{rev5}
\cite{rev6}
for some recent reviews and further references. 
For a detailed 
discussion of the 
gravitational collapse of radiation shells within a 
Vaidya metric and the related black hole and naked singularity geometries, 
we refer to
\cite{joshi}).

Basing 
on the work on spherical gravitational collapse done in
\cite{jd},
\cite{jg},
and 
\cite{gj},
we consider here 
the genericity of naked singularities forming in such a process. 
Given the data on the 
initial density and pressure profiles of the collapsing 
cloud, classes of solutions to Einstein's 
equations have been constructed which 
evolve to either a visible or a covered singularity, subject to the 
satisfaction of regularity 
and energy conditions. In other words, 
given the 
initial matter data on a spacelike surface from which the collapse 
of the massive matter cloud evolves, the rest of the free functions, 
such as the velocities of the collapsing shells as well as the
the classes of evolution, can be chosen 
as allowed by the Einstein equations. The latter 
take the collapse either to a black hole or a naked singularity in the final
state, depending on the choice made. The basic formalism here can be 
summarized as follows. 
The form of matter considered
is quite generic, which is any 
type I general matter field subject to an energy condition
\cite{he}.
In the case of a black hole developing as the endstate of collapse, 
the spacetime singularity is necessarily hidden behind the event
horizon of gravity. In contrast, 
a naked singularity develops when 
families of future directed non-spacelike trajectories come 
out from the singularity. These trajectories 
can in principle communicate information to 
faraway observers in the spacetime. The existence of such families confirms 
the formation of a naked singulzarity, 
as opposed to a black hole endstate.

The eventual singularity, which is the singularity curve 
produced by the collapsing matter, is what we study. 
We show here that 
the tangent to this curve at the central 
singularity at $r=0$ is related to the radially outgoing null geodesics 
from the singularity, 
if any. By determining 
the nature of the singularity curve and its relation to the initial 
data and the classes of collapse evolution, we are able to deduce whether 
the formation of any trapped surface during the collapse takes place before 
or after the singularity. It is this causal structure of the trapped region 
that determines the possible emergence or otherwise of non-spacelike paths 
from the singularity. That then settles the final outcome of the collapse in 
terms of either a black hole or naked singularity. Several familiar equations 
of state for which extensive collapse studies have been made, such as 
dust or matter with only tangential or radial pressures and others, are 
special cases of our consideration.

The spacetime geometry within the spherically symmetric collapsing 
cloud is described by the general metric in the comoving   
coordinates $(t,r,\theta,\phi)$ as 
\begin{equation}
ds^2=-e^{2\nu}dt^2+e^{2\psi}dr^2+R^2(t,r)d\Omega^2
\label{eq:metric}
\end{equation}
where $d\Omega^2$ is the line element on a two-sphere. The matter fields, 
considered by us, belong to a 
broad class, called {\it type I}, where the
energy-momentum tensor has one timelike and three spacelike eigenvectors.
This general class includes most physically reasonable matter, including dust, 
perfect fluids, massless scalar fields and such 
others. The stress-energy tensor for this class 
is given in a diagonal form: 
$T^t_t=-\rho, T^r_r=p_r, T^{\theta}_{\theta}=T^\phi_\phi=p_\theta$. 
Here 
$\rho$, $p_r$ and $p_\theta$ are the energy density, the radial pressure
and the tangential pressure respectively. We also take the matter field 
to satisfy the {\it weak energy condition}. This means that the energy density 
measured by any local observer must be non-negative  So, for any 
timelike vector $V^i$, one must have $T_{ik}V^iV^k\ge0$. The latter 
amounts to taking $\rho\ge0,\; \rho+p_r\ge0,\; \rho+p_\theta\ge0$. 
Now, for the metric (\ref{eq:metric}), Einstein's equations take the form 
(in the units $8\pi G=c=1$):

\begin{equation}
\rho=\frac{F'}{R^2R'}; \,\, p_r=-\frac{\dot{F}}{R^2\dot{R}},
\label{eq:ein1}
\end{equation}
\begin{equation}
\nu'=\frac{2(p_\theta-p_r)}{\rho+p_r}\frac{R'}{R}-\frac{p_r'}{\rho+p_r}
\label{eq:ein2}
\end{equation}
\begin{equation}
-2\dot{R}'+R'\frac{\dot{G}}{G}+\dot{R}\frac{H'}{H}=0
\label{eq:ein3}
\end{equation}
\begin{equation}
G-H=1-\frac{F}{R}
\label{eq:ein4},
\end{equation}
where we have defined $G(t,r)=e^{-2\psi}(R')^2$ and 
$H(t,r)=e^{-2\nu}(\dot{R})^2$.

The arbitrary function $F=F(t,r)$ has the interpretation of
the mass function for the cloud, giving the total mass in a shell of 
comoving radius $r$. The energy condition implies that  $F\ge0$. In order 
to preserve 
regularity at the initial epoch, we need to take $F(t_i,0)=0$, i.e. the 
mass function must vanish at the centre of the cloud. As seen from 
equation (4), 
there is a density singularity
in the spacetime at $R=0$, and one at $R'=0$. However, the latter one 
is due to shell-crossings and can be possibly removed  
\cite{clarke} through a suitable extension of the spacetime. 
In any case, we are interested here only in the shell-focusing 
singularity at $R=0$. This is a physical singularity where all the matter 
shells collapse to a zero physical radius. We can use the scaling 
freedom available for the radial coordinate $r$ to write $R=r$ at the 
initial epoch $t=t_i$ from where the collapse commences. Introducing the
function $v(t,r)$ by the relation
\begin{equation}
R(t,r)=rv(t,r),
\label{eq:R}
\end{equation}
we have $v(t_i,r)=1$ and $v(t_s(r),r)=0$. The collapse situation is now 
characterized by the condition $\dot{v}<0$.
The time $t=t_s(r)$ corresponds to the shell-focusing singularity 
$R=0$ where all the matter shells collapse to a vanishing physical
radius.

From the standpoint of the dynamic evolution of the initial data,
prescribed at the initial epoch $t=t_i$, there are six arbitrary functions 
of the comoving shell-radius $r$ as given by,
$\nu(t_i,r)=\nu_0(r), \psi(t_i,r)=\psi_0(r), R(t_i,r)=r 
\rho(t_i,r)=\rho_0(r), p_r(t_i,r)=p_{r_0}(r), \ 
p_\theta(t_i,r)=p_{\theta_0}$.
We note that not all of the initial data above are mutually independent, 
because from equation (\ref{eq:ein2}) one gets a relation that 
gives the initial function $\nu_0(r)$ in terms of rest
of the other initial data functions.
%\begin{equation}
%\nu_0(r)=\int_0^r\left(\frac{2(p_{\theta_0}-p_{r_0})}{r(\rho_0+p_{r_0})}
%-\frac{p_{r_0}'}{\rho_0+p_{r_0}}\right)dr.
%\label{eq:nu0}
%\end{equation}
The initial pressures should have a physically 
reasonable behaviour at the centre $(r=0)$ in that the pressure gradients 
should vanish there, 
i.e. $p_{r_0}'(0)=p_{\theta_0}'(0)=0$, and also the difference 
between the 
radial and the 
tangential pressures needs to 
vanish at the centre, 
i.e. $p_{r_0}(0)- p_{\theta_0}(0)=0$. These conditions are necessary
to ensure the regularity 
of the 
initial data at the centre of the cloud. 
It is then evident from equation (9) that $\nu_0(r)$ has the 
form
\begin{equation}
\nu_0(r)=r^2g(r),
\label{eq:nu0form}
\end{equation} 
where $g(r)$ is at least a $C^1$ function of $r$ for $r=0$, and at least 
a $C^2$ function for $r>0$. We 
see that there are five total field equations with seven 
unknowns, $\rho$, $p_r$, $p_{\theta}$, $\psi$, $\nu$, $R$, and $F$.  Thus we 
have the freedom to choose two free functions. In general, their selection, 
subject to the weak energy condition and the given initial data for 
collapse at the starting
surface, determines the matter distribution and the 
metric of the spacetime, and thus leads to a particular time evolution 
of the initial data.

Let us construct the classes of solutions to Einstein's equations
which give the collapse evolution from the given initial data, 
Consider first the general class of mass functions $F(t,r)$ for the collapsing
cloud. These are given as
\begin{equation}
F(t,r)=r^3{\cal M} (r,v),
\label{eq:mass}
\end{equation}
$\cal M$ being any regular and suitably differentiable general 
function without further restrictions. It turns out from equations given
below that the 
regularity and finiteness of the density profile at the initial epoch 
$t=t_i$ requires $F$ to go as $r^3$ close to the centre. It follows 
that the form of $F$ above is not really any special choice, but is in the
general class of mass functions consistent with the collapse regularity 
conditions. Equations (\ref{eq:ein1}) 
yield 
\begin{eqnarray}
\rho=\frac{3{\cal M}+r\left[{\cal M}_{,r}+{\cal M}_{,v}v'\right]}{v^2(v+rv')};
\,\, p_r=-\frac{{\cal M}_{,v}}{v^2}.
\label{eq:rho} 
\end{eqnarray}
Thus the regular density distribution at the initial epoch is given by 
$\rho_0(r)=3{\cal M}(r,1)+r{\cal M}(r,1)_{,r}$. 
It is evident in general that, as
$v\rightarrow 0$, $\rho\rightarrow\infty$ and $p_r\rightarrow\infty$, i.e. 
both the 
density and the radial pressure blow up at the singularity. One can, in fact, 
show the following. Given any regular initial density
and pressure profiles for the matter cloud from which the collapse 
develops, there always exist velocity profiles for collapsing matter 
shells as well as classes of dynamical evolutions, as determined by the 
Einstein equations, which lead to as naked singularity or a black hole as 
the endstate of collapse, depending on the choice of the class..

We now provide classes of solutions to
Einstein's equations to this effect. Consider the class of velocity profiles 
as determined by the general function
\begin{equation}
\nu=A(t,R),
\label{eq:nu}
\end{equation}
where $A(t,R)$ is any arbitrary, suitably differentiable 
function of $t$ and $R$, with the initial 
constraint $A(t_i,R)=\nu_0(r)$.
Using  (\ref{eq:nu}) in  (\ref{eq:ein3}), we have 
\begin{equation}
G(t,r)=b(r)e^{2(A-\int A_{,t}dt)}.
\label{eq:G}
\end{equation}
Here $b(r)$ is another arbitrary function of $r$ which emerges on 
integrating the Einstein equations. A comparison
with dust models enables one to interpret $b(r)$ as the velocity function for
the shells. From equation (\ref{eq:nu0form}), the form 
of $A(t,R)$ is seen to be $A(t,R)=r^2g_1(r,v)$, where $g_1(r,v)$ is 
a suitably differentiable function and $g_1(r,1)=g(r)$.
Similarly, we have $A-\int A_{,t}dt=r^2g_2(r,v)$
and at the initial epoch $g_2(r,1)=g(r)$.
Using (\ref{eq:nu}) in (\ref{eq:ein2}), we obtain
\begin{equation}
2p_\theta=RA_{,R}(\rho+p_r)+2 p_r+\frac{Rp_r'}{R'}.
\label{eq:ptheta}
\end{equation}
In general, both the density and the radial pressure are known to blow up at 
the singularity. However, the above equation implies that the tangential 
pressure also does the same.

Writing
\begin{equation}
b(r)=1+r^2b_0(r)
\label{eq:veldist}
\end{equation}
and using  equations (\ref{eq:mass}),(\ref{eq:nu}) and (\ref{eq:G}) 
in (\ref{eq:ein4}), we are led to
\begin{equation}
\sqrt{R}\dot{R}=-e^{r^2g_1(r,v)}\sqrt{(1+r^2b_0)Re^{r^2g_2(r,v)}-R+r^3{\cal
M}}. 
\label{eq:collapse}
\end{equation}
Since we are considering a collapse situation, we have $\dot{R}<0$.
Defining a function $h(r,v)$ as
\begin{equation}
h(r,v)=\frac{e^{2r^2g_2(r,v)}-1}{r^2}=2g_2(r,v)+{\cal O}(r^2v^2)
\label{eq:h}
\end{equation}
and using (\ref{eq:h}) in (\ref{eq:collapse}), we obtain  
after some simplifications:
\begin{equation}
\sqrt{v}\dot{v}=-\sqrt{e^{2r^2(g_1+g_2)}vb_0+e^{2r^2g_1}\left(vh(r,v)
+{\cal M} (r,v)\right)}.
\label{eq:collapse1}
\end{equation}
Integrating the above equation we have
\begin{equation}
t(v,r)=\int_v^1\frac{\sqrt{v}dv}{\sqrt{e^{2r^2(g_1+g_2)}vb_0+e^{2r^2g_1}
\left(vh+{\cal M}\right)}}.
\label{eq:scurve1}
\end{equation}
Note that $r$ is treated as a constant in the above integration. 
Expanding $t(v,r)$ around the centre, we have
\begin{equation} 
t(v,r)=t(v,0)+r{\cal X}(v)+{\cal O}(r^2),
\label{eq:scurve2}
\end{equation}
where the function ${\cal X}(v)$ is given by,
\begin{equation}
{\cal X}(v)=-\frac{1}{2}\int_v^1dv\frac{\sqrt{v}(b_1v+vh_1(v)+{\cal M}_1(v))}
{\left(b_{00}v+vh_0(v)+{\cal M}_0(v)\right)^{\frac{3}{2}}},
\label{eq:tangent1}
\end{equation}
with, $b_{00}=b_0(0), {\cal M}_0(v)={\cal M}(0,v), 
h_0=h(0,v), b_1=b_0'(0), M_1(v)={\cal M}_{,r}(0,v) and h_1=h_{,r}(0,v)$.

From the above, the time when the central singularity develops is 
given by
\begin{equation}
t_{s_0}=\int_0^1\frac{\sqrt{v}dv}{\sqrt{b_{00}v+vh_0(v)+{\cal M}_0(v)}}
\label{eq:scurve3}.
\end{equation}
The time for other shells to reach the singularity  
can be given by the expansion
\begin{equation}
t_s(r)=t_{s_0}+r{\cal X}(0)+{\cal O}(r^2).
\label{eq:scurve4}
\end{equation}
It is now clear that the value of ${\cal X}(0)$ depends 
on the functions $b_0,\cal M$ and $h$, which in turn depend on 
the initial data at $t=t_i$ and on the dynamical variable $v$
that evolves in time from a value $v=1$ at the initial epoch to
$v=0$ at the singularity. 
Thus, a given set of initial matter distribution and the 
dynamical profiles including the velocity of shells completely determine 
the tangent at the centre to the singularity curve.
Equations 
(17-19), 
lead to 
\begin{equation}
\sqrt{v}v'={\cal X} (v)\sqrt{b_{00}v+vh_0(v)+{\cal M}_0(v)}+{\cal O}(r^2).
\label{eq:scurve5}
\end{equation}

The endstate of collapse, in the form of either a black hole or 
a naked singularity, is determined by the causal behaviour of
the apparent horizon, which is the boundary of 
trapped surfaces forming due to collapse, and is given by $R=F$. 
If the neighbourhood 
of the centre gets trapped prior to the epoch of singularity, then it 
is covered and a black hole results, Otherwise, it would be a naked 
singularity with the 
non-spacelike future directed trajectories escaping from it.
It is then to be determined if there are any families of future directed
non-spacelike paths emerging from the singularity. To see this as well as
to examine the nature of the central singularity at $R=0$, 
$r=0$, one has to consider the equation 
\begin{equation} 
\frac{dt}{dr}=e^{\psi-\nu}
\end{equation}
for outgoing radial null geodesics. 
For a further discussion on this, we refer to 
\cite{jd},
\cite{jg},
and 
\cite{gj}.
The main result that follows from this is that if the quantity 
${\cal X}(0)$ is positive, one indeed gets  
radially outgoing null geodesics coming out from the central singularity, which
then has to be naked. 
However, if ${\cal X}(0)$ is negative, we have a black hole solution, since 
there are 
no such trajectories coming out. If ${\cal X}(0)$ vanishes, then we have 
to take into account the next higher order non-zero term in the 
singularity curve $t_s(r)$, and a similar analysis can be 
carried out.

In the above discussion, the functions $h$ and ${\cal M}$ have been
expanded in $r$ around $r=0$ and only the first-order terms
have been retained. The key point here is that the functions
chosen for the classes of collapse evolution from the given initial
data are such that the eventual singularity curve $t_s(r)$ is 
expandable at the centre $r=0$, at least to first order. The
well-known and extensively studied example of dust collapse 
(see e.g.   
\cite{jd93} 
and references therein)
does satisfy this feature. It thus forms a special sub-class of the above
classes of solutions for spherical collapse. Various other collapse models
studied earlier, such as collapse with a non-zero tangential pressure
and others also satisfy such a behaviour of the singularity curve. 
For a detailed discussion of the differentiability conditions on
the functions involved and of the classes of models that satisfy these,
we refer to the papers cited above.
The treatment given here applies to general 
classes of functions with the minimum differentiability conditions. 
However, in discussions of collapse these are sometimes assumed to be
analytic and expandable with respect to $r^{2}$ with the argument 
that such smooth functions are physically more 
relevant. Such assumptions are in fact due to computational
convenience, though one can mathematically exploit the freedom of definition 
and choice available. The formalism, outlined above, 
would of course work for such smooth functions, which constitute a special 
case of what has been considered. Consequently, one can have a smooth and 
differentiable singularity curve.

We thus see how the initial data 
determine the black hole and naked singularity phases as collapse 
endstates in terms of the available free functions. This happens because the 
quantity ${\cal X}(0)$ is determined by these initial 
and dynamical profiles, as given by (\ref{eq:tangent1}). It is clear 
that, given any regular density and pressure profiles for
the matter cloud from which the collapse develops, one can always 
choose velocity profiles such that the endstate of the collapse would
be either a naked singularity or a black hole, and vice-versa.
It is interesting to note here that physical agencies, such as the
spacetime shear within a dynamically collapsing cloud, could naturally
give rise to such phases in gravitational collapse 
\cite{jdm}.
In other words, such physical factors can naturally delay the
formation of any apparent horizon and trapped surfaces.

As stated above, we have worked here with the {\it type I} matter fields. 
This is a rather general form of matter which includes
practically all known physically reasonable fields such as dust,
prefect fluids, massless scalar fields and so on. However, note that we 
have assumed no explicit equation of state
relating the density and pressure variables. Assuming a
specific equation of state will fully close the system, all the
concerned functions being then determined as a result of the  development 
of the system from the initial data. We presently 
have little idea about the kind of equation of
state that the matter should follow. This is 
particularly true when the matter is close to the collapse
endstate, where we are dealing with ultra-high densities
and pressures. One might as well be allowed to 
freely choose the property of the matter fields, as done above. 
It is nevertheless 
important to make the following point. 
The analysis given above does include several well-known classes of collapse 
models and equations of state. We had a choice of two free functions 
available and 
constructed general classes of collapse evolution using the mass function
$F(r,v)$ and the metric function $\nu$. Suppose now that, in addition to 
Einstein's equations, an equation of state of the form $p_r=f(\rho)$ (or 
alternatively, $p_\theta=g(\rho)$) is given. Then, as evident from 
equation (\ref{eq:ein1}), 
there would be a constraint on the otherwise arbitrary function 
$\cal M$, specifying the required class, provided a solution to the constraint 
equation exists. Then the value of $p_\theta$ (or $p_r$) gets determined 
in terms of 
$\rho(t,r)$ from equation (\ref{eq:ein2}), and the 
analysis for the black hole and naked singularity phases 
goes through as above.

For example, in the case of collapsing dust models, we have
\begin{equation} 
p_r=p_\theta=0
\end{equation} 
and the constraint equation yields 
\begin{equation} 
{\cal M} (r,v)={\cal M} (r).
\end{equation} 
Next, for collapse with a constant (or zero)
radial pressure, but with an allowed variable tangential pressure, 
the constraint equation leads to
\begin{equation}
{\cal M}(r,v)=f(r)-kv^3,
\end{equation}
where, $k$ is the value of the constant radial pressure. The tangential
pressure will then be given by ,
\begin{equation}
p_\theta=k+\frac{1}{2}A_{,R}R(\rho+k).
\end{equation}
Along with these two well-known models, the analysis works for
any other model in which any one of the pressures is specified by an 
equation of state and which permits a solution to the constraint equation 
on $\cal M$. This provides some insight into how these black hole and
naked singularity phases come about as collapse endstates.

The above results provide us with information on the
genericity aspects of naked singularities developing as the final states
of spherical collapse. Consider a specific collapse evolution for 
which the quantity ${\cal X}(0)$ is positive, leading to the formation of a 
naked singularity as the endstate. As noted earlier, this
value is decided by the initial density and pressure profiles. 
The initial velocity profiles 
for the collapsing shells are controlled by 
the function $b_0(r)$ and the
collapse evolutions ${\cal M} (r,v)$ and $h(r,v)$, which in turn depend on 
the initial data at $t=t_i$ and the dynamical variable $v$. Hence,
by continuity, any arbitrarily small variation in any of these parameters
will continue to yield a value for ${\cal X}(0)$ which is again positive. 
Thus if a collapse evolution 
creates a naked singularity, any sufficiently small variation in the
initial matter and velocity profiles for the collapsing shell, or 
small variations
in the dynamical evolutions functions will still preserve the final outcome
of a naked singularity. In this sense the naked singularities of 
spherical collapse turn out to be completely generic.

We paraphrase the basic result discussed in this article. 
Given a starting set of regular data 
in terms of the density and pressure profiles at the initial 
epoch from which the collapse develops, there are sets of dynamical  
evolution, i.e. classes of solutions to the Einstein equations, which
can evolve the given initial data to produce either a black hole or a
naked singularity as the collapse endstate. The exact outcome 
depends on the choice of the rest of the free functions available, Indeed, 
the space of initial data can be divided into two 
distinct subspaces: one containing those that evolve into black holes and 
another containing those that lead to 
naked singularities. The latter are generic in the
sense described above. In a way this brings out the stability of 
these collapse outcomes with respect to perturbations in the initial data set
and in the choice of collapsing matter fields.

\section{5. Concluding remarks}

In this article we have discussed genericity aspects of spacetime 
singularities. It is seen that in Einstein gravity they occur generically,
whether covered within event horizons or as visible to external
observers. This is a consequence of the singularity theorems which make crucial
use of the Raychaudhuri equation. It is possible that a future quantum theory 
of gravity may remove 
or resolve the final singularity of collapse or the initial one 
in cosmology. In such a 
scenario, what matter physically are the regions of ultra-strong 
gravity and spacetime curvature that develop as a result
of the accompanying  dynamical gravitational processes. This is true even if 
the very final singularity  
is removed through quantum effects. The following physical picture then
emerges. 
Dynamical gravitational processes proceed and evolve to create 
ultra-strong gravity regions in the universe. Once these form, strong 
curvature and quantum effects both 
come into their own in these regions. Quantum gravity then takes over and is 
likely to resolve the final singularity. 
Particularly interesting is the case when the singularities
of collapse are visible. In such a situation, quantum gravity effects,
taking place in those ultra-strong gravity regions, will in principle be 
accessible and observable to external observers. The consequences are 
likely to be intriguing.
\cite{gjs}.
\bigskip

{\it Acknowledgment:} I would like to express my deep gratitude
to the late  A. K. Raychaudhuri for his keen interest and extremely relevant
comments on our work on gravitational collapse and spacetime singularities.
It was a great pleasure knowing and discussing with him, and
to benefit from his penetrating insights.

\end{document}